\documentclass[12pt]{article}
\usepackage{amsmath,amssymb,amsthm,amsxtra,overpic,bbm,bm,epsfig,ulem}
\usepackage{color}

\usepackage{cite}

\usepackage{hyperref}
\usepackage{url}

\def\thefootnote{\fnsymbol{footnote}}

\begin{document}

\vspace{0.3cm}

\begin{center}
{\large\bf Parameter degeneracy and information loss in inverse flavor
mapping for high-energy astrophysical neutrinos} \\[0.6cm]
			
{\small\bf Zi-Qiang~Chen}$^{a,b,c,}$\footnote{chenziqiang22@mails.ucas.ac.cn},
{\small\bf Zhi-zhong~Xing}$^{d,e,}$\footnote{xingzz@ihep.ac.cn},
{\small\bf Ye-Ling~Zhou}$^{a,}$\footnote{zhouyeling@ucas.ac.cn}
\\[-2mm]
\end{center}

\vspace*{0.10cm}

\centerline{$^{a}$\small\it School of Fundamental Physics and Mathematical Sciences,}
\centerline{\small\it Hangzhou Institute for Advanced Study, UCAS, Hangzhou 310024,
China}
\centerline{$^{b}$\small\it Institute of Theoretical Physics, Chinese Academy of
Sciences, Beijing 100190, China}
\centerline{$^{c}$\small\it University of Chinese Academy of Sciences, Beijing 100049,
China}
\centerline{$^{d}$\small\it Institute of High Energy Physics, Chinese Academy of
Sciences,100049, Beijing, China}
\centerline{$^{e}$\small\it Center of High Energy Physics, Peking University,100871,
Beijing, China}

\vspace*{1cm}
		
\begin{abstract}
Given a high-energy astrophysical neutrino flux, its source flavor
ratios $\eta = \{\eta^{}_e, \eta^{}_\mu, \eta^{}_\tau\}$ are
correlated with the ones
$f = \{f^{}_e, f^{}_\mu, f^{}_\tau\}$ measured at a neutrino telescope
via $f = P \eta$, where the elements of $P$ consist of four lepton
flavor mixing parameters. But realistic {\it inverse} flavor
mapping $\eta = P^{-1} f$ encounters unavoidable parameter degeneracy
and information loss, especially in or near the $\det P = 0$ limit
allowed by current neutrino oscillation data. In this work we explore
the parameter correlation condition for $\det P = 0$ including the 
$\mu$-$\tau$ reflection symmetry case, formulate the corresponding 
constraints on $f^{}_e$ and $f^{}_\mu$ versus $\eta^{}_e$
and $\eta^{}_\mu$, and illustrate to what extent the source
flavor distribution can be mapped by using the recent IceCube all-sky
neutrino flux data ranging from 5 TeV to 10 PeV under the assumption
that the relevant sources have a common flavor composition. 
\end{abstract}

\newpage

\def\thefootnote{\arabic{footnote}}
\setcounter{footnote}{0}

\section{Introduction}
\label{1}

As one of the cosmic messengers, very high-energy neutrinos
offer a direct probe of the hadronic processes in a remote
astrophysical source or environment. The electrical neutrality
and weak interaction properties of such an energetic neutrino flux
assure that it can not only travel over astronomical distances
without losing directional information but also escape the regions
that are opaque to electromagnetic radiation, although its energy-
and distance-dependent information will be lost when being detected
at a neutrino telescope due to unavoidable quantum
decoherence~\cite{Learned:2000sw,Halzen:2002pg,Winter:2012xq}. So
far the IceCube telescope has observed high-energy extraterrestrial
neutrinos~\cite{IceCube:2013low}, established the Glashow
resonance~\cite{IceCube:2021rpz}, and even measured the flavor
composition of an all-sky neutrino flux ranging from $5 ~{\rm TeV}$
to $10 ~{\rm PeV}$~\cite{IceCube:2025ole}. In comparison, the
recently running KM3NeT telescope has reported an exceptional
high-energy astrophysical neutrino event~\cite{KM3NeT:2025npi}.
A few newly proposed neutrino telescope projects are also
underway~\cite{Baikal-GVD:2018isr,TRIDENT:2022hql,Chen:2026jdx}.

The flavor composition of a high-energy astrophysical neutrino flux
observed at a terrestrial neutrino telescope depends upon the production
processes at the source, the evolution of secondary particles around
the source, and the effects of flavor oscillations during neutrino
propagation~\cite{Beacom:2003nh,Xing:2006xd,Xing:2006uk,Becker:2007sv,
Bustamante:2015waa,Bustamante:2019sdb,Song:2020nfh}. Given
that the total flux of three-flavor astrophysical
neutrinos and antineutrinos is conserved, one may describe their
source flavor distribution by using the normalized column vector
$\eta \equiv \{\eta^{}_e , \eta^{}_\mu , \eta^{}_\tau\}^T$ with
$\eta^{}_e + \eta^{}_\mu + \eta^{}_\tau = 1$. The corresponding
telescope flavor composition can be analogously described by
$f \equiv \{f^{}_e, f^{}_\mu, f^{}_\tau\}^T$ with
$f^{}_e + f^{}_\mu + f^{}_\tau = 1$. So far several {\it idealized}
sources of high-energy astrophysical neutrinos have been conjectured
and studied as typical
examples~\cite{Learned:1994wg,Anchordoqui:2003vc,Lipari:2007su,
Kashti:2005qa}, and among them the case of $\eta = \{1/3, 2/3, 0\}^T$
is expected to be most likely because it can naturally originate from
the cascade decays of charged pions produced from high-energy
proton-proton and proton-gamma collisions.
For some practical reasons, however, the flavor composition of a
{\it realistic} high-energy astrophysical neutrino flux may be more or
less different from any previously conjectured ones~\cite{Hummer:2010ai}.

Given the astronomical baseline between an astrophysical source and a
neutrino telescope, variations of the oscillation phases over the source
energy spectrum and the experimental energy bins may cause the energy-
and distance-dependent interference terms of neutrino oscillations to
average to zero. In this case, it is a highly simplified neutrino
oscillation probability matrix $P$ that bridges the gap between $\eta$
and $f$:
\begin{eqnarray}
\left[\begin{matrix}
f^{}_e \cr f^{}_\mu \cr f^{}_\tau \end{matrix}\right]
= \left(\begin{matrix}
P^{}_{e e} & P^{}_{e \mu} & P^{}_{e \tau} \cr
P^{}_{\mu e} & P^{}_{\mu \mu} & P^{}_{\mu \tau} \cr
P^{}_{\tau e} & P^{}_{\tau \mu} & P^{}_{\tau \tau}
\end{matrix}\right)
\left[\begin{matrix} \eta^{}_e \cr \eta^{}_\mu \cr
\eta^{}_\tau \end{matrix}\right] \; ,
\label{1}
\end{eqnarray}
where $P^{}_{\alpha\beta}$ (for $\alpha, \beta = e, \mu, \tau$) are
expressed in terms of nine elements of the $3 \times 3$ unitary lepton
flavor mixing matrix $U$ as follows:
\begin{eqnarray}
P^{}_{\alpha\beta} = P^{}_{\beta\alpha} = \sum^3_{i=1}
\left(|U^{}_{\alpha i}|^2 \hspace{0.05cm} |U^{}_{\beta i}|^2\right) \; .
\label{2}
\end{eqnarray}
The Particle-Data-Group-advocated Euler-like parametrization of $U$ sets
$U^{}_{e 1} = \cos\theta^{}_{12} \cos\theta^{}_{13}$,
$U^{}_{e 2} = \sin\theta^{}_{12} \cos\theta^{}_{13}$,
$U^{}_{e 3} = \sin\theta^{}_{13} e^{-{\rm i}\delta}$,
$U^{}_{\mu 3} = \cos\theta^{}_{13} \sin\theta^{}_{23}$ and
$U^{}_{\tau 3} = \cos\theta^{}_{13} \cos\theta^{}_{23}$ with
$\delta$ being the Dirac phase responsible for CP violation in neutrino
oscillations, and the other four elements of $U$ can be derived by using
the normalization and orthogonality conditions of
$U$~\cite{ParticleDataGroup:2026}. As the elements in each row or column
of $P$ are positive and their sum is always equal to one, the parameter
space of $f^{}_e$, $f^{}_\mu$ and $f^{}_\tau$ mapped from that of
$\eta^{}_e$, $\eta^{}_\mu$ and $\eta^{}_\tau$ via $f = P \eta$ is naturally
expected to converge rather than diverge.

But the primary goal of a neutrino telescope is actually to
solve {\it the inverse problem}: how to infer the source flavor
composition $\eta$ from the observed components of $f$ by means of
the inverse mapping formula $\eta = P^{-1} f$, which will be divergent
if $\det P \to 0$ (e.g., in or near the $\mu$-$\tau$ flavor symmetry
limit~\cite{Lai:2009ke,Fu:2012zr,Fu:2014isa,Xing:2025xte}). Given the
very fact that the currently available neutrino oscillation
data {\it do} support an approximate $\mu$-$\tau$ reflection symmetry
associated with the three active neutrino flavors,
one needs to find a workable solution to the above inverse problem so
as to model-independently map the source flavors of a high-energy
astrophysical neutrino flux with the help of the terrestrial
telescope experiments.

The present work is intended to report some new and significant
progress that we have made in tracing the original flavor composition
of high-energy astrophysical neutrinos. We find that there exists
unavoidable parameter degeneracy behind $\det P = 0$, in which the
$\mu$-$\tau$ reflection symmetry is just a special case.
To reduce this kind of parameter degeneracy, $\theta^{}_{23}$ and
$\delta$ have to be determined to an unprecedented degree of accuracy.
Furthermore, we derive very novel constraint equations for $f^{}_e$ and
$f^{}_\mu$ versus $\eta^{}_e$ and $\eta^{}_\mu$ in the $\det P = 0$
limit, and illustrate to what extent the source flavor distribution
can be mapped by taking account of the recent IceCube all-sky neutrino
flux data ranging from 5 TeV to 10 PeV under the assumption that the 
relevant sources have a common flavor composition. The results
obtained in this work are expected to find useful applications in 
neutrino telescope phenomenology.

\section{Parameter degeneracy}
\label{2}

\begin{figure}[t]
\centering
\includegraphics[width=1\textwidth]{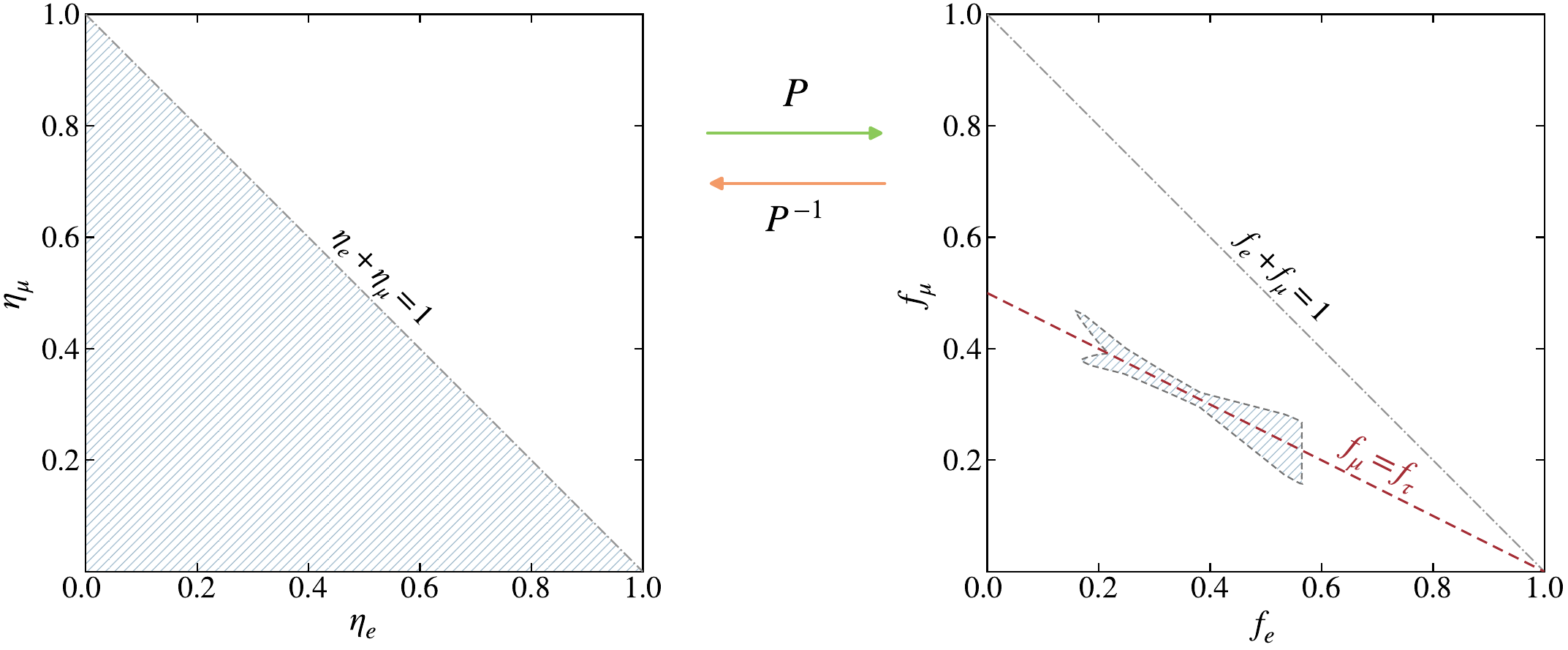}
\vspace{-0.9cm}
\caption{The flavor composition diagram for an arbitrary
high-energy astrophysical neutrino source with
$\eta^{}_\alpha \in \left[0, 1\right]$ (for $\alpha = e, \mu, \tau$)
and its {\it dress-like} counterpart mapped via $f = P \eta$ at a
neutrino telescope, where the $3\sigma$ ranges of $\theta^{}_{12}$,
$\theta^{}_{13}$, $\theta^{}_{23}$ and $\delta$ in the normal mass
ordering case are input~\cite{Esteban:2024eli}, and the red dashed
line results from the $\mu$-$\tau$ reflection symmetry.}
\label{fig1}
\end{figure}

We begin with a general $f = P \eta$ mapping
by allowing the source flavor components $\eta^{}_e$, $\eta^{}_\mu$
and $\eta^{}_\tau$ to vary in their respective $\left[0, 1\right]$
intervals but keeping $\eta^{}_e + \eta^{}_\mu + \eta^{}_\tau = 1$
unchanged. Inputting the $3\sigma$ ranges of $\theta^{}_{12}$,
$\theta^{}_{13}$, $\theta^{}_{23}$ and $\delta$ reported in
the NuFIT version 6.1 for the normal neutrino mass
ordering~\cite{Esteban:2024eli}, we plot the trivial two-dimensional
source flavor diagram and the nontrivial $P$-bridged mapping diagram
for the flavor composition at a neutrino telescope in Fig.~\ref{fig1}.
A very interesting finding of this numerical illustration is that the
strongly constrained region of $f^{}_e$, $f^{}_\mu$ and $f^{}_\tau$
is roughly {\it dress-like} and symmetric about the red dashed axis
fixed by $f^{}_\mu = f^{}_\tau$, a direct consequence of the
$\mu$-$\tau$ reflection symmetry of $U$ (i.e., $\theta^{}_{23} = \pi/4$
and $\delta = 3\pi/2$~\cite{Harrison:2002et,Xing:2006ms,Zhou:2014sya,
Xing:2015fdg,Xing:2022uax}) which is independent of the source flavor
ratios.

Fig.~\ref{fig1} is a very clear indication that the inverse flavor
mapping via $\eta = P^{-1} f$ is in general divergent and hence requires
sufficiently good degrees of accuracy for the flavor mixing parameters
of $U$ and the observed flavor components of $f$. In particular, this
inverse problem will be unsolvable in the $\det P = 0$ limit, a
possibility which cannot be ruled out by current neutrino oscillation
data. But we find that a partial solution to such a special inverse
problem is actually possible, as can be seen later on in section~\ref{3}.

\begin{figure}[t]
\centering
\includegraphics[width=0.625\textwidth]{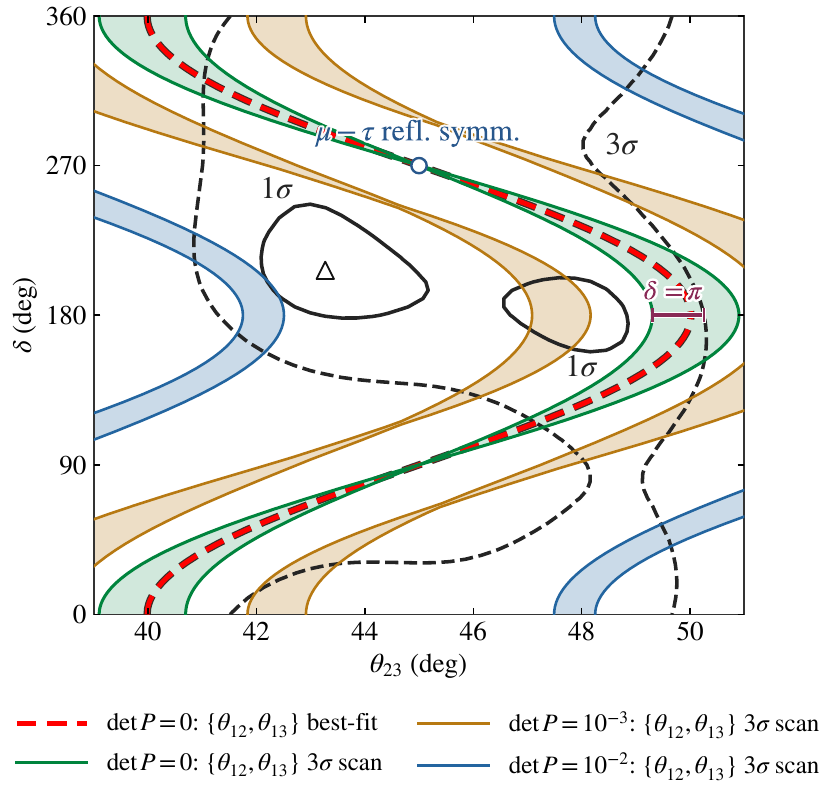}
\vspace{-0.3cm}
\caption{The best-fit (black triangle), $1\sigma$-allowed (black
solid closed curves) and $3\sigma$-allowed (black dashed curves)
values of $\delta$ and $\theta^{}_{23}$ reported by the NuFIT
group~\cite{Esteban:2024eli}, together
with their correlation bands constrained by Eq.~(\ref{3}) in three
specific cases: $\det P = 0$ (red dashed curve and light green
region with the white dot marking the $\mu$-$\tau$ reflection
symmetry point and the purple line segment marking the
range of $\theta^{}_{23}$ with respect to $\delta = \pi$ inside
the NuFIT $3\sigma$ region), $\det P = 10^{-3}$ (light brown region) 
and $\det P = 10^{-2}$ (light blue region).}
\label{fig2}
\end{figure}

Here let us work out a general parameter correlation condition for
$\det P = 0$. A lengthy but straightforward calculation of $\det P$
leads us to
\begin{eqnarray}
\det P \hspace{-0.2cm} & = & \hspace{-0.2cm}
\left| \begin{matrix}
P^{}_{ee} & P^{}_{e\mu} & P^{}_{e\tau} \cr
P^{}_{e\mu} & P^{}_{\mu\mu} & P^{}_{\mu\tau} \cr
P^{}_{e\tau} & P^{}_{\mu\tau} & P^{}_{\tau\tau} \end{matrix}
\right|
\nonumber \\
\hspace{-0.2cm} & = & \hspace{-0.2cm}
\left| \begin{matrix}
P^{}_{ee} - P^{}_{e\tau} & P^{}_{e\mu} - P^{}_{e\tau} \cr
P^{}_{e\mu} - P^{}_{\mu\tau} & P^{}_{\mu\mu} - P^{}_{\mu\tau} \cr
\end{matrix}
\right|
\nonumber \\
\hspace{-0.2cm} & = & \hspace{-0.2cm}
\left(P^{}_{ee} - P^{}_{e\tau}\right) \left(P^{}_{\mu\mu} - P^{}_{\mu\tau}\right)
- \left(P^{}_{e\mu} - P^{}_{e\tau}\right) \left(P^{}_{e\mu} - P^{}_{\mu\tau}\right)
\nonumber \\
\hspace{-0.2cm} & = & \hspace{-0.2cm}
\Big[ |U^{}_{e 1}|^2 \left(|U^{}_{\mu 2}|^2
|U^{}_{\tau 3}|^2 - |U^{}_{\mu 3}|^2
|U^{}_{\tau 2}|^2\right)
+ |U^{}_{e 2}|^2 \left(|U^{}_{\mu 3}|^2
|U^{}_{\tau 1}|^2 - |U^{}_{\mu 1}|^2
|U^{}_{\tau 3}|^2\right)
\nonumber \\
\hspace{-0.2cm} & & \hspace{-0.2cm} +
\hspace{0.1cm} |U^{}_{e 3}|^2 \left(|U^{}_{\mu 1}|^2
|U^{}_{\tau 2}|^2 - |U^{}_{\mu 2}|^2
|U^{}_{\tau 1}|^2\right) \Big]^2
\nonumber \\
\hspace{-0.2cm} & = & \hspace{-0.2cm}
\left[\cos 2\theta^{}_{12} \cos 2\theta^{}_{13}
\cos 2\theta^{}_{23} - \frac{1}{2} \sin 2\theta^{}_{12}
\sin\theta^{}_{13} \left(1 - 3\sin^2\theta^{}_{13}\right)
\sin 2\theta^{}_{23} \cos\delta\right]^2 \; , \hspace{0.6cm}
\label{3}
\end{eqnarray}
implying that $\det P \geq 0$ must hold. As a result,
$\det P = 0$ gives rise to an interesting correlation between the
CP-violating phase and three flavor mixing angles:
\begin{eqnarray}
\cot 2\theta^{}_{23} = \frac{1}{2} \tan 2\theta^{}_{12}
\hspace{0.06cm}
\frac{\sin\theta^{}_{13} \left(1 - 3\sin^2\theta^{}_{13}\right)}
{\cos 2\theta^{}_{13}} \hspace{0.02cm} \cos\delta \; ,
\label{4}
\end{eqnarray}
which is sensitive to the octant of $\theta^{}_{23}$ and the
quadrant of $\delta$. But it is worth pointing out that Eq.~(\ref{4})
can also be derived from $\det |U|^2 = 0$ thanks to the relation
$\det P = \left(\det |U|^2\right)^2$, where $|U|^2$ stands 
for a $3\times 3$ matrix whose nine elements are 
$|U^{}_{\alpha\beta}|^2$ (for $\alpha, \beta = e, \mu, 
\tau$)~\cite{Fu:2014isa}. Today's neutrino oscillation data have
excluded the possibility of $\theta^{}_{13} = 0$, but
$\theta^{}_{23} = \pi/4$ and $\delta = 3\pi/2$ are surely allowed
at the $\gtrsim 2\sigma$ confidence level~\cite{Esteban:2024eli}.
So the $\mu$-$\tau$ reflection symmetry results in
$\det P = 0$ and obstructs a complete mapping of $\eta$ from $f$ via
$\eta = P^{-1} f$. But Eq.~(\ref{4})
actually determines a narrow band in the plane of $\theta^{}_{23}$
and $\delta$ after the best-fit and $3\sigma$-allowed values of
$\theta^{}_{12}$ and $\theta^{}_{13}$ from the NuFIT version 6.1
are input, as numerically illustrated in Fig.~\ref{fig2}
(the red dashed curve and light green region), from which
parameter degeneracy can be easily observed.

Of course, unavoidable parameter degeneracy may also emerge for
a given nonzero value of $\det P$. In Fig.~\ref{fig2} we have also
illustrated the allowed ranges of $\theta^{}_{23}$ and $\delta$ in the
typical case of $\det P = 10^{-3}$ (light brown region) or
$\det P = 10^{-2}$ (light blue region). One may easily see that
the $10^{-3} \lesssim \det P \lesssim 10^{-2}$ region is more or less
favored by today's neutrino oscillation data. To reduce this
parameter degeneracy, it is necessary to determine $\theta^{}_{23}$ and
$\delta$ to an unprecedented degree of accuracy in the future 
long-baseline neutrino oscillation experiments~\cite{DUNE:2015lol,
Hyper-KamiokandeProto-:2015xww}.

One should also keep in mind that the recent JUNO 
measurement~\cite{JUNO:2025gmd}, which has reported an unprecedented 
precision value of $\theta^{}_{12}$ but has not been included into 
the NuFIT version 6.1 used in our numerical calculations, may help 
narrow the parameter space shown in Fig.~\ref{fig2}. 

\section{Information loss}
\label{3}

We proceed to study how much information about the telescope and source
flavor compositions can be obtained in the $\det P = 0$ limit. Given
that it is usually more difficult to measure the tau flavor at a neutrino
telescope, let us rewrite Eq.~(\ref{1}) with the help of
$f^{}_\tau = 1 - f^{}_e - f^{}_\mu$ and
$\eta^{}_\tau = 1 - \eta^{}_e - \eta^{}_\mu$ in the following way:
\begin{eqnarray}
\left[\begin{matrix}
f^{}_e - P^{}_{e\tau} \cr f^{}_\mu - P^{}_{\mu\tau} \end{matrix}\right]
= \left(\begin{matrix}
P^{}_{ee} - P^{}_{e\tau} & P^{}_{e\mu} - P^{}_{e\tau} \cr
P^{}_{e\mu} - P^{}_{\mu\tau} & P^{}_{\mu\mu} - P^{}_{\mu\tau}
\end{matrix}\right)
\left[\begin{matrix} \eta^{}_e \cr \eta^{}_\mu \end{matrix}\right] \; ,
\label{5}
\end{eqnarray}
in which the $2 \times 2$ coefficient matrix has the same
determinant as $P$, as can be seen from Eq.~(\ref{3}).
Following Cramer's rule, we express the two solutions to
Eq.~(\ref{5}) as
\begin{eqnarray}
\eta^{}_e \cdot \det P \hspace{-0.2cm} & = & \hspace{-0.2cm}
\left| \begin{matrix}
f^{}_e - P^{}_{e\tau} & P^{}_{e\mu} - P^{}_{e\tau} \cr
f^{}_\mu - P^{}_{\mu\tau} & P^{}_{\mu\mu} - P^{}_{\mu\tau} \end{matrix}
\right| \; ,
\nonumber \\
\eta^{}_\mu \cdot \det P \hspace{-0.2cm} & = & \hspace{-0.2cm}
\left| \begin{matrix}
P^{}_{ee} - P^{}_{e\tau} & \hspace{0.03cm} f^{}_e - P^{}_{e\tau} \cr
P^{}_{e\mu} - P^{}_{\mu\tau} & \hspace{0.03cm} f^{}_\mu - P^{}_{\mu\tau}
\end{matrix}
\right| \; . \hspace{1.1cm}
\label{6}
\end{eqnarray}
So $\det P = 0$ allows us to obtain a novel linear constraint equation
for $f^{}_e$ and $f^{}_\mu$ from Eq.~(\ref{6}) and then a similar
constraint formula for $\eta^{}_e$ and $\eta^{}_\mu$ with the help of
Eq.~(\ref{5}):
\begin{eqnarray}
f^{}_\mu \hspace{-0.2cm} & = & \hspace{-0.2cm}
\frac{f^{}_e \left(P^{}_{\mu\mu} - P^{}_{\mu\tau}\right) +
P^{}_{e\mu} P^{}_{\mu\tau} - P^{}_{e\tau} P^{}_{\mu\mu}}
{P^{}_{e\mu} - P^{}_{e\tau}} \; ,
\nonumber \\
\hspace{-0.2cm} & = & \hspace{-0.2cm}
\frac{\displaystyle \sum^3_{i=1} |U^{}_{ei}|^2 \left[\left(3 |U^{}_{\mu i}|^2
- 1\right) f^{}_e + |U^{}_{ei}|^2 - |U^{}_{\mu i}|^2\right]}
{\displaystyle \sum^3_{i=1} |U^{}_{ei}|^2 \left(3 |U^{}_{ei}|^2 - 1\right)}
\nonumber \\
\hspace{-0.2cm} & = & \hspace{-0.2cm}
\frac{1 - f^{}_e}{2} + \frac{\left(3 f^{}_e - 1\right) \cos 2\theta^{}_{23}}
{2 \left(1 - 2 \tan^2\theta^{}_{13}\right)} \; ,
\nonumber \\ \vspace{0.5cm} \nonumber \\
\eta^{}_\mu \hspace{-0.2cm} & = & \hspace{-0.2cm}
\frac{\eta^{}_e \left(P^{}_{e\tau} - P^{}_{ee}\right) +
f^{}_e - P^{}_{e\tau}}{P^{}_{e\mu} - P^{}_{e\tau}}
\nonumber \\
\hspace{-0.2cm} & = & \hspace{-0.2cm}
\frac{\displaystyle \sum^3_{i=1} |U^{}_{ei}|^2 \left[f^{}_e -
\eta^{}_e \left(|U^{}_{ei}|^2 - |U^{}_{\tau i}|^2\right) - |U^{}_{\tau i}|^2
\right]}
{\displaystyle \sum^3_{i=1} |U^{}_{ei}|^2 \left(|U^{}_{\mu i}|^2 -
|U^{}_{\tau i}|^2\right)}
\nonumber \\
\hspace{-0.2cm} & = & \hspace{-0.2cm}
\frac{1 - \eta^{}_e}{2} - \frac{1 - 2 \tan^2\theta^{}_{13}}
{2 \cos 2\theta^{}_{23}} \left[\eta^{}_e -
\frac{4 f^{}_e - \sin^2 2\theta^{}_{12} \cos^4\theta^{}_{13}
- \sin^2 2\theta^{}_{13}}{3 \cos^2 2\theta^{}_{12} \cos^4\theta^{}_{13}
+ \left(1 - 3 \sin^2\theta^{}_{13}\right)^2}\right] \; , \hspace{0.4cm}
\label{7}
\end{eqnarray}
which are not directly dependent on the CP-violating phase $\delta$.
Note that the
term in the square bracket of $\eta^{}_\mu$ vanishes when $\det P = 0$
originates from the $\mu$-$\tau$ reflection symmetry, giving rise to
\begin{eqnarray}
\eta^{}_e = \frac{4 f^{}_e - \sin^2 2\theta^{}_{12} \cos^4\theta^{}_{13}
- \sin^2 2\theta^{}_{13}}{3 \cos^2 2\theta^{}_{12} \cos^4\theta^{}_{13}
+ \left(1 - 3 \sin^2\theta^{}_{13}\right)^2} \; ,
\label{8}
\end{eqnarray}
a result which has been achieved in Ref.~\cite{Xing:2025xte}.
In this more restrictive case with $f^{}_\mu = \left(1 - f^{}_e\right)/2$,
it is the combination $\eta^{}_\mu + \eta^{}_\tau = 1 - \eta^{}_e$ (instead
of the individual source flavor ratios $\eta^{}_\mu$ and $\eta^{}_\tau$)
that can be determined from the neutrino telescope experiments.

A physical interpretation of Eq.~(\ref{7}) is as follows. If $\det P = 0$
happens to hold, it implies that the four lepton flavor mixing parameters are
intrinsically correlated with one another as already shown in Eq.~(\ref{4}).
This special parameter correlation can {\it mathematically} lead us to the
above constraints on $f^{}_e$ and $f^{}_\mu$ versus $\eta^{}_e$ and
$\eta^{}_\mu$, which reflect the existence of unavoidable information
loss. In other words, it is impossible to separately infer the values of
$\eta^{}_e$ and $\eta^{}_\mu$ from those of $f^{}_e$ and $f^{}_\mu$
in the $\det P = 0$ limit, but it is possible to achieve a linear correlation
between $\eta^{}_e$ and $\eta^{}_\mu$ for the source flavor distribution
in this parameter degeneracy case. It is especially interesting to see that
$\det P = 0$ even dictates the telescope flavor ratios $f^{}_e$ and $f^{}_\mu$
to satisfy a very simple linear correlation as given in Eq.~(\ref{7}). Such a
novel prediction surely deserves an experimental test at a neutrino telescope
in the foreseeable future, provided the possibility of $\det P = 0$ remains
viable after more accurate data are accumulated from the forthcoming precision
measurements of various long-baseline neutrino oscillations.

\begin{figure}[h!]
\centering
\includegraphics[width=0.88\textwidth]{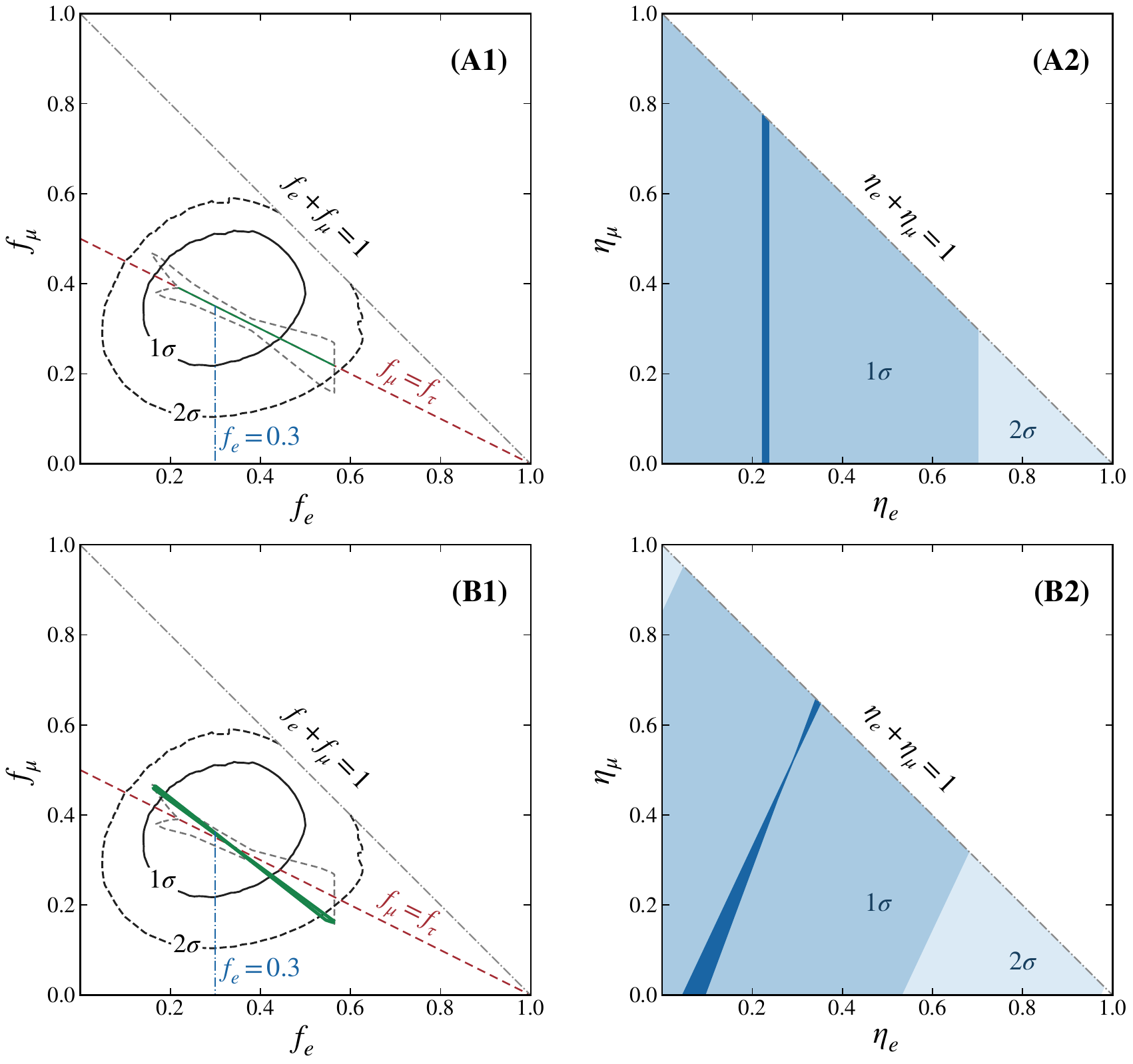}
\vspace{-0.2cm}
\caption{A linear correlation between $f^{}_e$ (or $\eta^{}_e$) and
$f^{}_\mu$ (or $\eta^{}_\mu$) as constrained in the $\det P = 0$ limit
with $\delta = 3\pi/2$ in (A1) and (A2) or $\delta = \pi$ in (B1) and 
(B2). The green line in (A1) or the green band in (B1) shows the 
allowed telescope flavor ratios located in the general {\it dress-like} 
region obtained in Fig.~\ref{fig1}. The blue and light blue regions in 
(A2) and (B2) exhibit the respective source flavor ratios which are 
compatible with the $1\sigma$ and $2\sigma$ contours of the recent 
IceCube telescope data~\cite{IceCube:2025ole}. The vertical dark blue 
band in (A2) and the dark blue oblique region in (B2) are inferred 
respectively from (A1) and (B1) with the same typical input 
$f^{}_e = 0.30$.}
\label{fig3}
\end{figure}

In view of Fig.~\ref{fig2}, which is plotted with the help of the data
extracted from a global analysis of today's available neutrino oscillation
measurements, we find that it is certainly too early to rule out the
possibility of $\det P = 0$ at the $3\sigma$ confidence level. In this 
situation let us illustrate information loss --- to what extent the source flavor
distribution can be mapped by adopting the recent IceCube all-sky neutrino
flux data ranging from 5 TeV to 10 PeV~\cite{IceCube:2025ole} and assuming the
relevant sources to have a common flavor composition. Two benchmark examples
are considered in Fig.~\ref{fig3}: (A) $\delta = 3\pi/2$ (maximal CP
violation) and (B) $\delta = \pi$ (CP conservation).
In both cases $\theta^{}_{12}$ and $\theta^{}_{13}$ are allowed to vary
in their respective $3\sigma$ ranges~\cite{Esteban:2024eli}.
As for $\delta = 3\pi/2$, Eq.~(\ref{4}) leads us to $\theta^{}_{23}
= \pi/4$ and hence the exact $\mu$-$\tau$ reflection symmetry limit
corresponding to the white dot in Fig.~\ref{fig2}. In the $\delta = \pi$
case, however, Eq.~(\ref{4}) tells us that $\theta^{}_{23}$ is located in a
parameter space resulting from the $3\sigma$ uncertainties of $\theta^{}_{12}$
and $\theta^{}_{13}$, corresponding to the purple line segment in Fig.~\ref{fig2}.
That is why the linear correlation between $f^{}_e$ and $f^{}_\mu$ in
Eq.~(\ref{7}) determines the green line in Fig.~\ref{fig3} (A1) or the green band
in Fig.~\ref{fig3} (B1) for the telescope flavor ratios, which must lie in the
general {\it dress-like} region obtained in Fig.~\ref{fig1}. Then we introduce
the $1\sigma$ and $2\sigma$ contours of the IceCube data into Fig.~\ref{fig3}
(A1) and (B1) in order to further constrain the allowed values of $f^{}_e$ and
$f^{}_\mu$ in the $\det P = 0$ limit, from which the source flavor distribution
can be consequently inferred in the blue and light blue regions of
Fig.~\ref{fig3} (A2) and (B2). It is obvious that the $2\sigma$ contour imposes
almost no bound on the source flavor ratios $\eta^{}_e$ and $\eta^{}_\mu$,
whereas the $1\sigma$ contour sets a seeable upper bound on $\eta^{}_e$
and a feeble upper limit on $\eta^{}_\mu$.

Note that the vertical dark blue band in Fig.~\ref{fig3} (A2) is obtained 
from Fig.~\ref{fig3} (A1) by typically taking $f^{}_e = 0.30$ (and hence 
$f^{}_\mu = 0.35$). In this case we arrive at the source flavor ratio
$\eta^{}_e \simeq 0.23$, but $\eta^{}_\mu$ itself is unconstrained as a
result of the $\mu$-$\tau$ reflection symmetry. In comparison, the dark blue
oblique region of $\eta^{}_e$ and $\eta^{}_\mu$ in Fig.~\ref{fig3} (B2) is
achieved from Fig.~\ref{fig3} (B1) with $\delta = \pi$ and $f^{}_e = 0.30$, 
but here $f^{}_\mu$ is allowed to take a set of different values around
$f^{}_\mu \simeq 0.36$ from the vertical slice of the green band 
corresponding to $f^{}_e = 0.30$. So the coordinates of $\eta^{}_e$ and 
$\eta^{}_\mu$ may vary roughly from $\left(0.08, 0.0\right)$
to $\left(0.35, 0.65\right)$ in Fig.~\ref{fig3} (B2) in the $\det P = 0$ limit.
With the improvement of precision and accuracy associated with the future
IceCube experiment~\cite{IceCube-Gen2:2025wru}, we hope to see whether the
possibility of $\det P = 0$ can be finally ruled out or not.

\section{Summary}
\label{4}

We have studied the inverse flavor mapping problem which is extremely 
important and relevant to neutrino telescope physics, in order to 
model-independently trace the original flavor composition of high-energy 
astrophysical neutrinos and probe the hadronic processes in the remote
source environment. An interesting finding is that there exists 
unavoidable parameter degeneracy behind $\det P = 0$, where the neutrino 
oscillation probability matrix $P$ bridges the gap between the source
flavor distribution $\eta = \{\eta^{}_e, \eta^{}_\mu, \eta^{}_\tau\}$
and the telescope one $f = \{f^{}_e, f^{}_\mu, f^{}_\tau\}$ via
$f = P \eta$ or $\eta = P^{-1} f$. The lepton flavor mixing parameters
$\theta^{}_{23}$ and $\delta$, which are crucially related to an underlying 
$\mu$-$\tau$ reflection flavor symmetry, need to be determined to an 
unprecedented degree of accuracy so as to reduce the above parameter degeneracy.
We have also derived very novel constraint equations for $f^{}_e$ and
$f^{}_\mu$ versus $\eta^{}_e$ and $\eta^{}_\mu$ in the $\det P = 0$
limit, and illustrated to what extent the source flavor distribution
can be mapped by taking advantage of the recent IceCube all-sky neutrino
flux data ranging from 5 TeV to 10 PeV under the reasonable assumption
that the relevant sources have a common flavor composition. 

We expect that this work will be an indispensable addition to
today's neutrino phenomenology and applicable to tomorrow's neutrino
telescope experiments. It is especially worth remarking that the 
next-generation precision neutrino oscillation experiments will help 
a lot for the success of the neutrino telescopes in astrophysics.

\section*{Acknowledgements}

The authors are indebted to Y.F. Li and S. Zhou for useful discussions.
This work was supported in part by the National Natural Science Foundation
of China under Grant No. 12535007 (ZZX and YLZ) and Grant No. 12547104
(YLZ), by the Scientific and Technological Innovation Program of the
Institute of High Energy Physics under Grant No. E55457U2 (ZZX), and by
the Zhejiang Provincial Natural Science Foundation of China under Grant
No. LDQ24A050002 (YLZ).

\end{document}